\documentclass{article}
\usepackage{spconf,amsmath,graphicx, cite, threadcol}
\usepackage{caption} 
\let\OLDthebibliography\thebibliography
\renewcommand\thebibliography[1]{
  \OLDthebibliography{#1}
  \setlength{\parskip}{0pt}
  \setlength{\itemsep}{1pt plus 0.3ex}
}

\title{Transformer-Transducer: \\ End-to-End Speech Recognition with Self-Attention}

\name{
\begin{tabular}{c}
Ching-Feng Yeh $^{\star}$, Jay Mahadeokar $^{\star}$, Kaustubh Kalgaonkar, Yongqiang Wang, \\
Duc Le, Mahaveer Jain, Kjell Schubert, Christian Fuegen, Michael L. Seltzer
\end{tabular}
\thanks{$\star$ Equal contribution.}
}

\address{Facebook AI, USA}

\begin{document}
\ninept
\maketitle
\begin{abstract}
We explore options to use Transformer networks in neural transducer for end-to-end speech recognition. Transformer networks use self-attention for sequence modeling and comes with advantages in parallel computation and capturing contexts. We propose 1) using VGGNet with causal convolution to incorporate positional information and reduce frame rate for efficient inference 2) using truncated self-attention to enable streaming for Transformer and reduce computational complexity. All experiments are conducted on the public LibriSpeech corpus. The proposed Transformer-Transducer outperforms neural transducer with LSTM/BLSTM networks and achieved word error rates of 6.37 \% on the \texttt{test-clean} set and 15.30 \% on the \texttt{test-other} set, while remaining streamable, compact with 45.7M parameters for the entire system, and computationally efficient with complexity of $O(T)$, where $T$ is input sequence length. 
\end{abstract}
\section{Introduction}
\label{sec:intro}

There has been significant progress on automatic speech recognition (ASR) technologies over the past few years due to the adoption of deep neural networks \cite{dahl2011context}. Conventionally, speech recognition systems involve individual components for explicit modeling on different levels of signal transformation: acoustic models for audio to acoustic units, pronunciation model for acoustic units to words and language model for words to sentences. This framework is often referred to as the ``traditional" hybrid system. Conventionally, individual components in the hybrid system can be optimized separately. For example, CD-DNN-HMM \cite{dahl2011context} focuses on maximizing the likelihood between acoustic signals and acoustic models with frame-level alignments. For language modeling, both statistical n-gram models \cite{brown1992class} and more recently, neural-network-based models \cite{mikolov2010recurrent} aim to model purely the connection between word tokens. 

Hybrid systems achieved significant success \cite{xiong2016achieving} but also present challenges. For example, hybrid system requires more human intervention in the building process, including the design of acoustic units, the vocabulary, the pronunciation model and more. In addition, an accurate hybrid system often comes with the cost of higher computational complexity and memory consumption, thus increasing the difficulty of deploying hybrid systems in resource-limited scenarios such as on-device speech recognition. Given the challenges, the interests in end-to-end approaches for speech recognition have surged recently \cite{graves2012sequence, he2019streaming, rao2017exploring, mohamed2019transformers, graves2006connectionist, graves2014towards,  chan2016listen, dong2019self}. Different from hybrid systems, end-to-end approaches aim to model the transformation from audio signal to word tokens directly, therefore the model becomes simpler and requires less human intervention. In addition to the simplicity of training process, end-to-end systems also demonstrated promising recognition accuracy \cite{chan2016listen}. Among many end-to-end approaches, recurrent neural network transducer (RNN-T) \cite{graves2012sequence, he2019streaming} provides promising potential on footprint, accuracy and efficiency. In this work, we explore options for further improvements based on RNN-T.

Recurrent neural networks (RNNs) such as long-short term memory (LSTM) \cite{hochreiter1997long} networks are good at sequence modeling and widely adopted for speech recognition. RNNs rely on the recurrent connection from the previous state $\boldsymbol{h}_{t-1}$ to the current state $\boldsymbol{h}_t$ to propagate contextual information. This recurrent connection is effective but also presents challenges. For example, since $\boldsymbol{h}_t$ depends on $\boldsymbol{h}_{t-1}$, RNNs are difficult to compute in parallel. In addition, $\boldsymbol{h}_t$ is usually of fixed dimensions, which means all historical information is condensed into a fixed-length vector and makes capturing long contexts also difficult. The attention mechanism \cite{vaswani2017attention, chorowski2015attention} was introduced recently as an alternative for sequence modeling. Compared with RNNs, the attention mechanism is non-recurrent and can compute in parallel easily. In addition, the attention mechanism can "attend" to longer contexts explicitly. With the attention mechanism, the Transformer model \cite{vaswani2017attention} achieved state-of-the-art performance in many sequence-to-sequence tasks \cite{chorowski2015attention, devlin2018bert}. 

In this paper, we explore options to apply Transformer networks in the neural transducer framework. VGG networks \cite{simonyan2014very} with causal convolution are adopted to incorporate contextual information into the Transformer networks and reduce the frame rate for efficient inference. In addition, we use truncated self-attention to enable streaming inference and reduce computational complexity . 

\section{Neural Transducer (RNN-T)}
\label{sec:neural_transducer}

In nature, speech recognition is a sequence-to-sequence (audio-to-text) task in which the lengths of input and output sequences can vary. As an end-to-end approach, connectionist temporal classification (CTC) \cite{graves2006connectionist} was introduced before RNN-T to model such sequence-to-sequence transformation. Given input sequence $\mathbf{x}=(\boldsymbol{x}_1, \boldsymbol{x}_2, \cdots, \boldsymbol{x}_T)$, where $\boldsymbol{x}_t \in \mathcal{R}^d$ and $T$ is the input sequence length, output sequence $\mathbf{y}=({y}_1, {y}_2, \cdots, {y}_u)$, where ${y}_u \in \mathcal{Z}$ represent output symbols and $U$ is the output sequence length, CTC introduces an additional "blank" label $b$ and models the posterior probability of $\mathbf{y}$ given $\mathbf{x}$ by:
\begin{align}
    P(\mathbf{y} | \mathbf{x}) = \sum_{\scriptscriptstyle {\hat{\mathbf{y}} \in \boldsymbol{H}_{\rm ctc}(\mathbf{x}, \mathbf{y})}} \displaystyle\prod_{t=1}^{T} P(\hat{y}_{t} | \boldsymbol{x}_1 \cdots \boldsymbol{x}_T)
\end{align}
where $\hat{\mathbf{y}} = (\hat{y}_1, \hat{y}_2, \cdots \hat{y}_T) \in \boldsymbol{H}_{\rm ctc}(\mathbf{x}, \mathbf{y}) \subset {\{\mathcal{Z} \cup b\}}^{T}$ correspond to any possible paths such that after removing $b$ and repeated consecutive symbols of  $\hat{\mathbf{y}}$ yields $\mathbf{y}$.

\begin{figure}[hhh]
    \centering
    \includegraphics[scale=0.08]{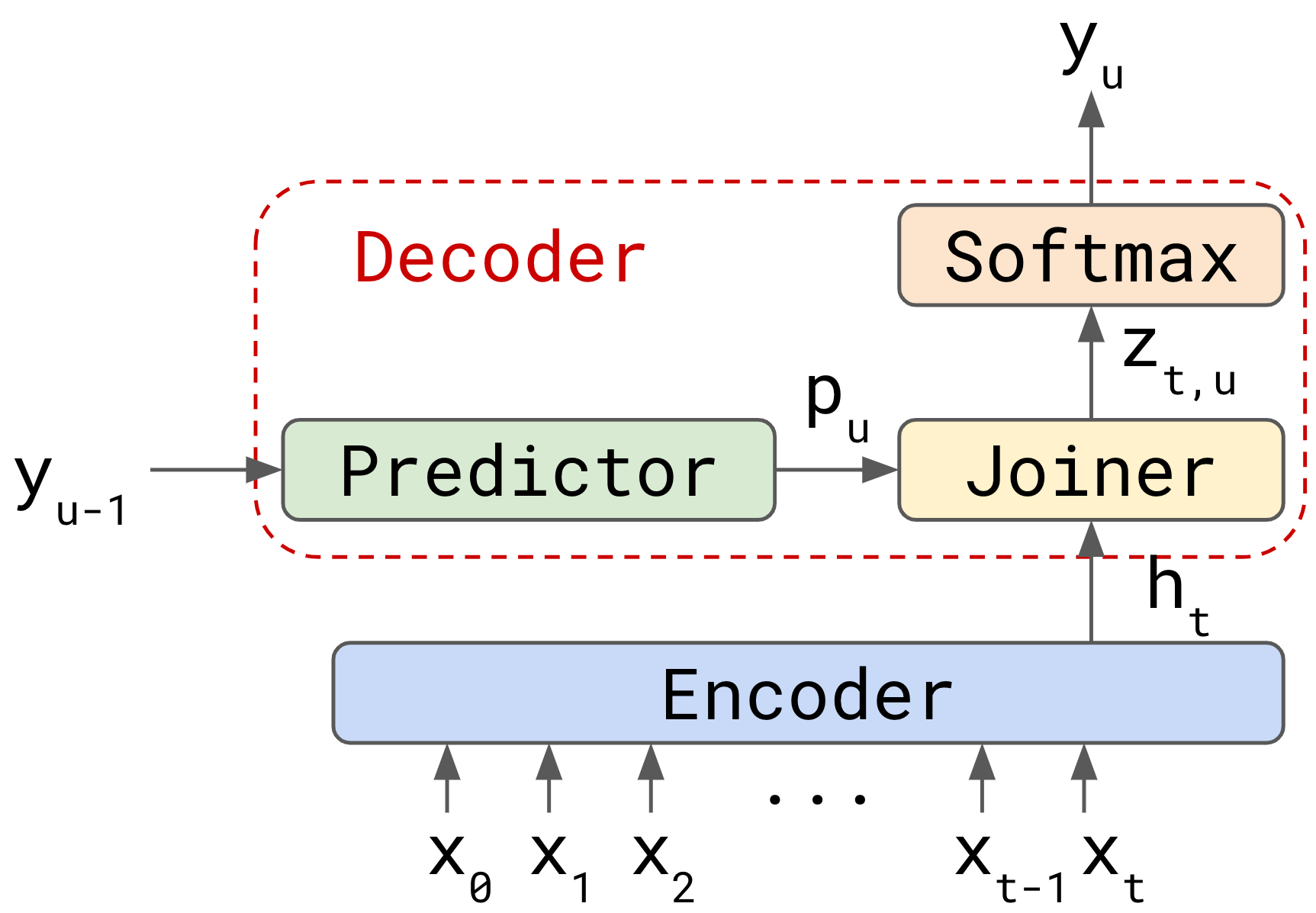}
    \caption{Neural Transducer.}
    \label{fig:neural_transducer}
\end{figure}

The formulation of CTC assumes that symbols in the output sequence are conditionally independent of one another given the input sequence. The RNN-T model improves upon CTC by making the output symbol distribution at each step dependent on the input sequence and previous non-blank output symbols in the history:
\begin{align}
    P(\mathbf{y} | \mathbf{x}) = \sum_{\scriptscriptstyle {\hat{\mathbf{y}} \in \boldsymbol{H}_{\rm rnnt}(\mathbf{x}, \mathbf{y})}} \displaystyle\prod_{i=1}^{T+U} P(\hat{y}_{i} | \boldsymbol{x}_1 \cdots \boldsymbol{x}_{t_{i}}, {y}_1 \cdots {y}_{u_{i-1}})
\end{align}
where $\hat{\mathbf{y}} = (\hat{y}_1, \hat{y}_2, \cdots \hat{y}_{T+U}) \in \boldsymbol{H}_{\rm rnnt}(\mathbf{x}, \mathbf{y}) \subset {\{\mathcal{Z} \cup b\}}^{T+U}$ correspond to any possible paths such that after removing $b$ and repeated consecutive symbols of  $\hat{\mathbf{y}}$ yields $\mathbf{y}$. By explicitly conditioning the current output on the history, RNN-T outperforms CTC when no external language model is present \cite{he2019streaming, rao2017exploring}. RNN-T can be implemented in the encoder-decoder framework, as illustrated in Fig. \ref{fig:neural_transducer}. The encoder encodes the input acoustic sequence $\mathbf{x}=(\boldsymbol{x}_1, \boldsymbol{x}_2, \cdots, \boldsymbol{x}_T)$ to $\mathbf{h}=(\boldsymbol{h}_1, \boldsymbol{h}_2, \cdots, \boldsymbol{h}_{T'})$ with potential subsampling $T' \leq T$. And the decoder contains a predictor to encode the previous non-blank output symbol ${y}_{u-1}$ for the logits $\boldsymbol{z}_{t,u}$ to condition on. It's worth noting that only when the most probable symbol $\hat{y}_{u}$ is non-blank the input to predictor $y_{u-1}$ will be updated, so that the conditioning encoding $\boldsymbol{p}_u$ only changes when non-blank output symbols are observed. From the illustration, we see that RNN-T incorporates a language model of output symbols internally in the decoder. 

There are many architectures that can be used as encoders and predictors. The functionality of these blocks is to take a sequence and find a higher-order representations. Recurrent neural networks (RNNs) such as LSTM \cite{hochreiter1997long} have been successfully used for such functionality. In this paper, we explore Transformer \cite{vaswani2017attention, chorowski2015attention} as an alternative for sequence encoding in RNN-T. Since Transformer is not recurrent in nature, we refer to the architecture illustrated in Fig. \ref{fig:neural_transducer} as simply "neural transducer" \cite{battenberg2017exploring} for the rest of the paper.

\section{Transformer}
\label{sec:transformer}

The attention mechanism \cite{bahdanau2014neural} is one of the core ideas of Transformer \cite{chorowski2015attention}. It was proposed to model correlation between contextual signals and produced state-of-the-art performance in many domains including machine translation \cite{chorowski2015attention} and natural language processing \cite{vaswani2017attention}. Similar to RNNs, attention mechanism aims to encode the input sequence to a higher-level representation by formulating the encoding function into the relationship between queries $Q$, keys $K$ and values $V$ and describing the similarities between them with:
\begin{align}
    Attention(Q, K, V) = Softmax(\frac{QK^T}{\sqrt{d_k}})V
\end{align}
where $Q \in \mathcal{R}^{T_q \times d_k}$, $K \in \mathcal{R}^{T_k \times d_k}$ and $V \in \mathcal{R}^{T_k \times d_v}$. This mechanism becomes "self-attention" when $Q = K = V = (\boldsymbol{x}_1, \cdots , \boldsymbol{x}_T)$. A self-attention block encodes the input $\mathbf{x}$ to  a higher-level representation $\mathbf{h}$, just like RNNs but without recurrence. Compared with RNNs where $\boldsymbol{h}_{t}$ depends on $\boldsymbol{h}_{t-1}$, self-attention has no recurrent connections between time steps in the encoding $\mathbf{h}$, therefore it can generate encoding efficiently in parallel. In addition, compared with RNNs where contexts are condensed into fixed-length states for the next time step to condition on, self-attention "pays attention" to all available contexts to better model the context within the input sequence.

\subsection{Multi-Head Self-Attention}
\label{ssec:multi_head_self_attention}

The attention mechanism can be further extended to multi-head attention, in which 1) dimensions of input sequences are split into multiple chunks with multiple projections 2) each chunk goes through independent attention mechanisms 3) encodings from each chunks are concatenated then projected to produce the output encodings, as described with:
\begin{multline}
    MultiHeadAttention(Q, K, V) = [\boldsymbol{e}_{\rm 1}, \cdots, \boldsymbol{e}_{\rm H}]W^{o} \\
    where\quad\boldsymbol{e}_{\rm i} = Attention(QW^{Q}_{i}, KW^{K}_{i}, VW^{V}_{i})
\end{multline}
where $H$ is the number of heads, $d_{in}$ is the dimension of input sequence, $d_{k} = d_{in} / H$, $\boldsymbol{e}_{\rm i}$ is the encoding generated by head $i$, $W^{o} \in \mathcal{R}^{d_{in} \times d_{in}}$, $W^{Q}_{i} \in \mathcal{R}^{d_{in} \times d_{k}}$, $W^{K}_{i} \in \mathcal{R}^{d_{in} \times d_{k}}$ and $W^{V}_{i} \in \mathcal{R}^{d_{in} \times d_{v}}$. Multi-head attention integrates encodings generated from multiple subspaces to higher-dimensional representations \cite{chorowski2015attention}.

\subsection{Transformer Encoder}
\label{ssec:transformer_encoder}

\begin{figure}[hhh]
    \centering
    \includegraphics[scale=0.10]{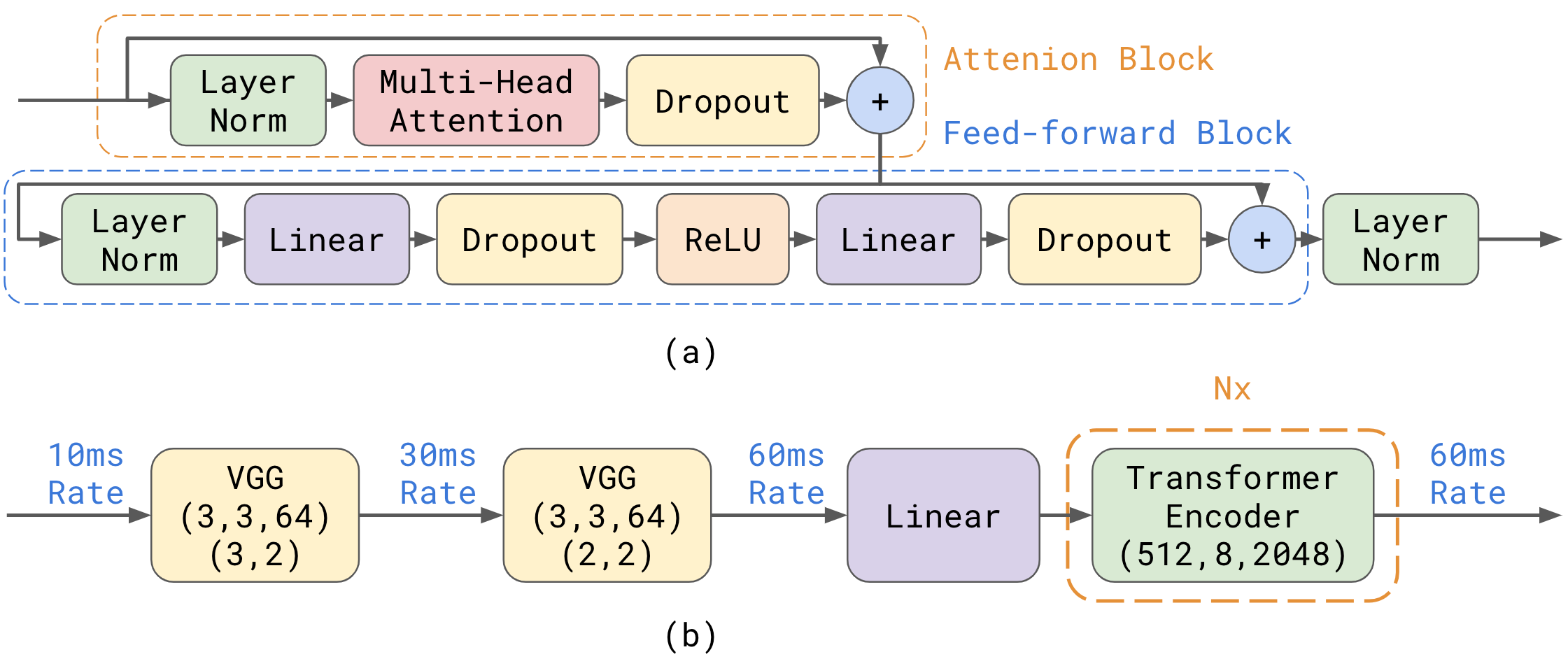}
    \caption{(a) Transformer Encoder (b) VGG-Transformer.}
    \label{fig:transformer_encoder_layer}
\end{figure}

The Transformer \cite{vaswani2017attention} is also a sequence-to-sequence model. The architecture of the Transformer encoder contains three main blocks: 1) attention block, 2) feed-forward block and 3) layer norm \cite{ba2016layer} as shown in Fig. \ref{fig:transformer_encoder_layer}(a).
The attention block contains the core multi-head self-attention component. The feed-forward block projects the input dimension $d_{in}$ to another feature space $d_{ff}$ and then back to $d_{in}$ (usually $d_{ff} \geq d_{in}$) for learning feature representation. The final layer normalization and other additional components including layer norm and dropout in the first two blocks are added to stabilize the model training and prevent overfitting. Furthermore, we use VGGNets to incorporate positional information into the Transformer as illustrated in Fig. \ref{fig:transformer_encoder_layer}(b). More details are given in section \ref{ssec:context_modeling}.

\section{Transformer-Transducer}
\label{sec:transformer_transducer}

Given the success of the Transformer, we explore the options of applying Transformer in neural transducer. For further improvement, we propose 1) using causal convolution for context modeling and frame rate reduction and 2) using truncated self-attention to reduce the computational complexity and enable streaming for Transformer.

\subsection{Context Modeling with Causal Convolution}
\label{ssec:context_modeling}

Transformer relies on multi-head self-attention to model the contextual information. However, attention mechanism is non-recurrent and non-convolutive, therefore risks losing the order or positional information in the input sequence \cite{gehring2017convolutional, chorowski2015attention}, which could harm the performance especially for the case of language modeling. To incorporate the positional information into Transformer, a simple way is adding positional encoding \cite{chorowski2015attention} but convolutional approaches \cite{mohamed2019transformers} demonstrated superior performance. In this paper we adopt the convolutional approach in \cite{mohamed2019transformers} with modification.

\begin{figure}[hhh]
    \centering
    \includegraphics[scale=0.07]{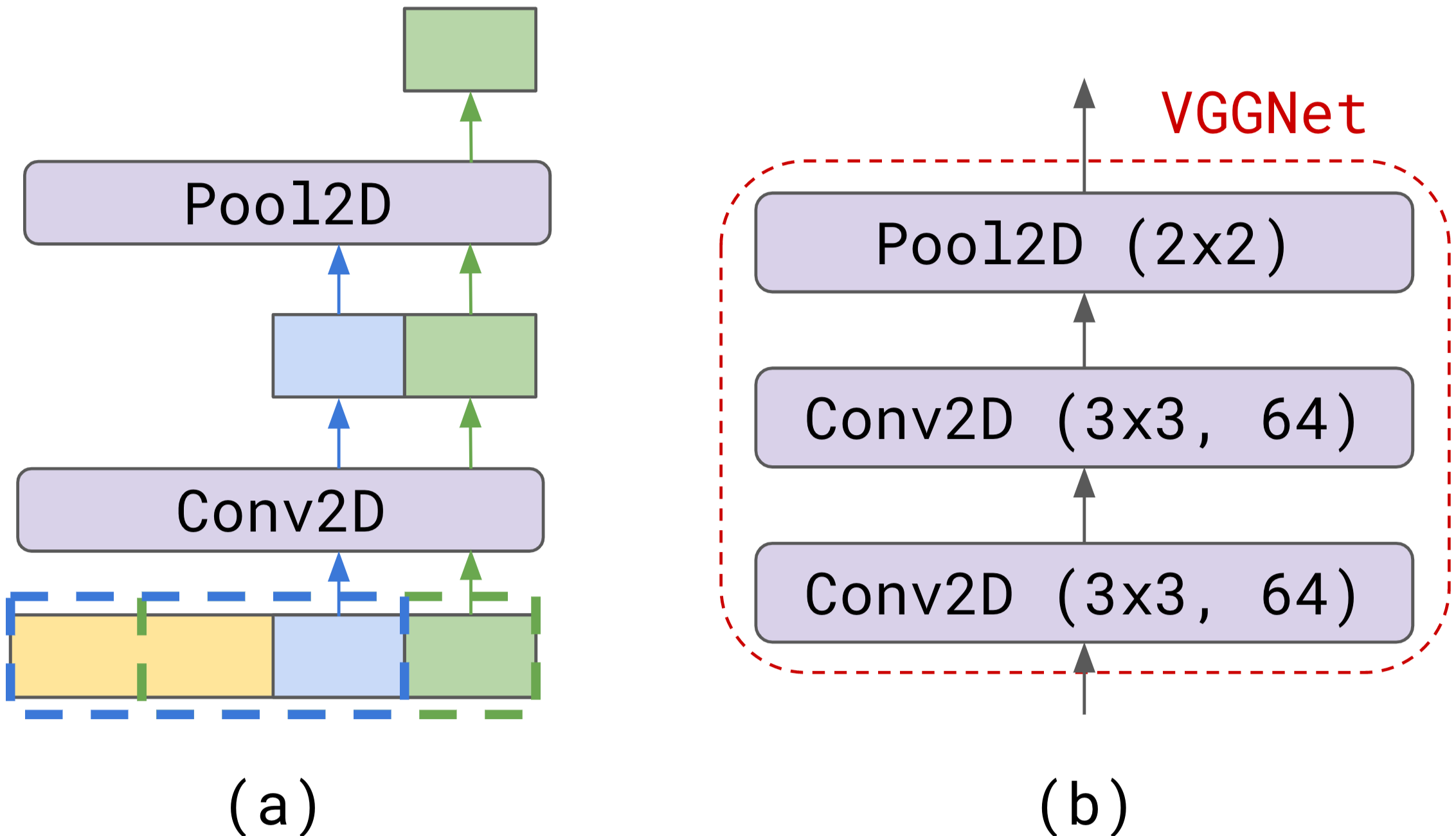}
    \caption{(a) Causal Convolution (b) VGGNet}
    \label{fig:causal_vgg_layer}
\end{figure}

Convolution networks model contexts by using kernels to convolve blocks of features. If we treat the input sequence (for example: acoustic features) as a two-dimensional image $X \in \mathcal{R}^{T \times D}$, in common practice for a $N \times K$ kernel the convolution would cover from $X(i - \frac{N-1}{2}, j - \frac{K-1}{2})$ to $X(i + \frac{N-1}{2}, j + \frac{K-1}{2})$ to produce the convolved output $Y(i, j)$. Therefore the convolution would need "future" information to generate the encoding for the current time step. For acoustic modeling this introduces additional look ahead and latency, but introducing future information is impractical for language modeling since the next symbol is unknown during inference.

To prevent future information from leaking into the computation at the current time step, we use causal convolution in which all contexts required are pushed to the history, as illustrated in Fig. \ref{fig:causal_vgg_layer}(a). With causal convolution, for a $N \times K$ kernel the convolution covers from $X(i - N + 1, j - \frac{K-1}{2})$ to $X(i, j + \frac{K-1}{2})$ to produce the convolved output $Y(i, j)$, therefore ensuring the convolution is purely "causal". Similar to \cite{mohamed2019transformers}, we also adopt the VGGNet \cite{simonyan2014very} structure, as illustrated in Fig. \ref{fig:causal_vgg_layer}(b), where two two-dimensional convolution layers are stacked sequentially followed by a two-dimensional max-pooling layer. We use layers of the causal VGGNet to incorporate positional information and propagate to the succeeding Transformer encoder layers. We refer to this network as "VGG-Transformer" and illustrate the architecture used for the encoder in neural transducer in Fig. \ref{sec:transformer}(b), where the first two VGGNet layers are used to incorporate positional information and reduce the frame rate for efficient inference, followed by a linear layer for dimension reduction and multiple Transformer encoder layers for generating higher-level representations. 

\subsection{Truncated Self-Attention}
\label{ssec:truncated_self_attention}

\begin{figure}[hhh]
    \centering
    \includegraphics[scale=0.07]{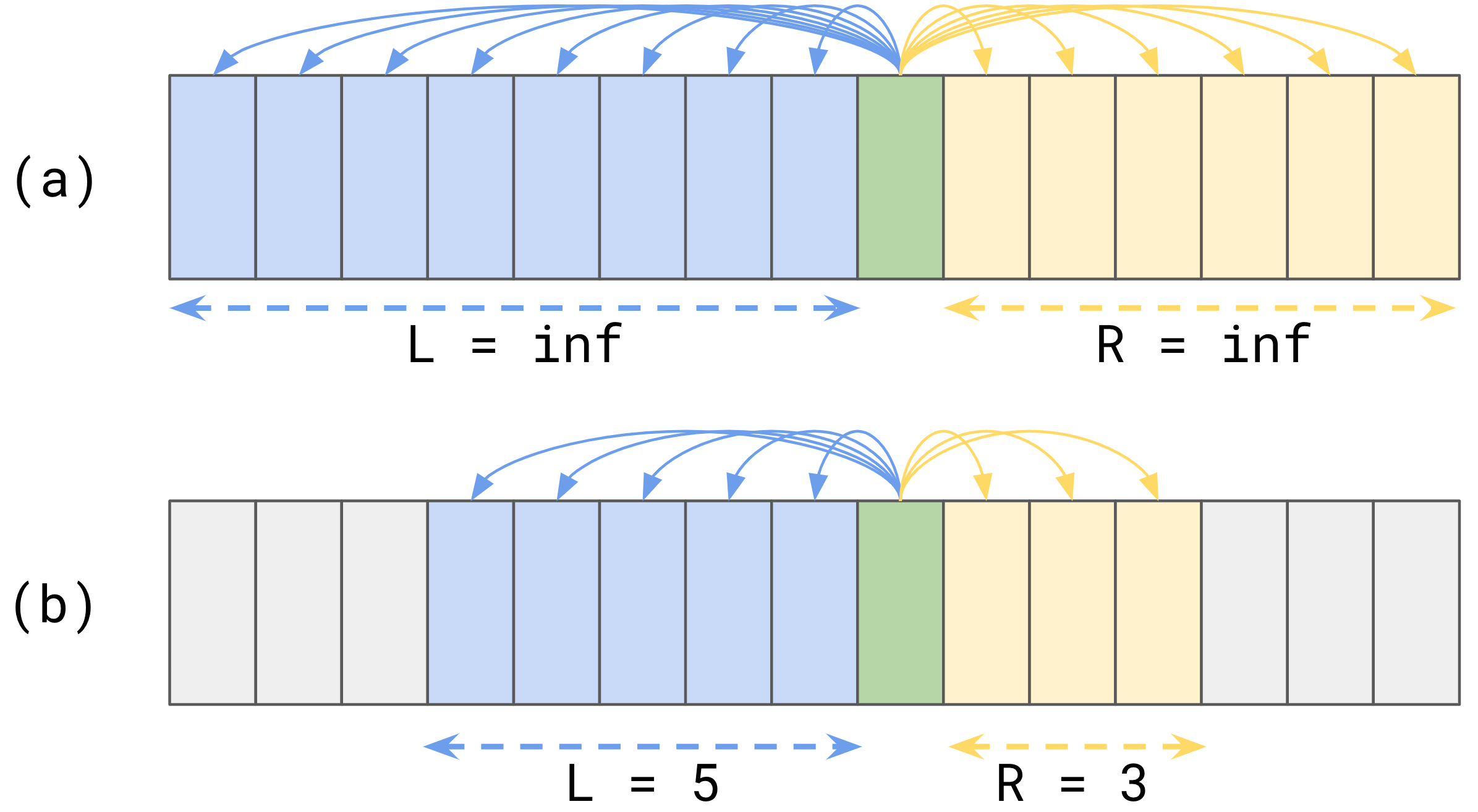}
    \caption{ Self-Attention: (a) Unlimited (b) Truncated.}
    \label{fig:truncated_self_attention}
\end{figure}

Unlimited self-attention attends to the whole input sequence and poses two issues: 1) streaming inference is disabled and 2) computational complexity is high. 
As illustrated in Fig. \ref{fig:truncated_self_attention}(a), for unlimited self-attention, the output $\boldsymbol{h}_{t}$ at time step $t$ depends on the entire input sequence $\mathbf{x}=(\boldsymbol{x}_1, \cdots, \boldsymbol{x}_T)$, meaning the inference can only begin after the final length $T$ is known. 
In addition, $\boldsymbol{h}_{t}$ depends on the similarity pairs $(\boldsymbol{x}_t, \boldsymbol{x}_1), (\boldsymbol{x}_t, \boldsymbol{x}_2) \cdots (\boldsymbol{x}_t, \boldsymbol{x}_T)$, giving complexity $O(T^{2})$ for computing $(\boldsymbol{h}_1, \cdots, \boldsymbol{h}_{T})$. 
These issues are critical for self-attention to work in scenarios demanding low-latency and low-computation such as on-device speech recognition \cite{he2019streaming}. 

To reduce both the latency and computational cost, we replace the unlimited self-attention by truncated self-attention, as illustrated in Fig. \ref{fig:truncated_self_attention}(b). Similar to time-delayed neural network (TDNN) \cite{waibel1995phoneme, peddinti2015time}, we limit the contexts available for self-attention so that output $\boldsymbol{h}_{t}$ at time $t$ only depends on $(\boldsymbol{x}_{t-L} \cdots \boldsymbol{x}_{t+R})$. Compared with unlimited self-attention, truncated self-attention is both streamable and computationally efficient. The look-ahead is the right context $R$ and the computational complexity reduces from $O(T^{2}$ to $O(T)$. However, it also comes with potential performance degradation and is investigated further in experiments.

\section{Experiments}
\label{sec:experiments}

\subsection{Corpus and Setup}
\label{ssec:dataset}

We use the publicly-available, widely-used LibriSpeech corpus \cite{panayotov2015librispeech} for experiments. LibriSpeech comes with 960 hours of read speech data for training, and 4 sets \texttt{\{dev, test\}-\{clean,other\}} for fine-tuning and evaluations. The \texttt{clean} sets contain high quality utterances where as the \texttt{other} sets are more acoustically challenging. We use \texttt{dev-\{clean,other\}} sets to fine-tune parameters for beam search and report results on \texttt{test-\{clean,other\}} results. We extract 80-dimensional log Mel-filter bank features every 10ms as acoustic features and normalize them with global mean computed from the training set. We also apply SpecAugment \cite{park2019specaugment} with policy "LD" for data distortion. A sentence piece model \cite{kudo2018sentencepiece} with 256 symbols is trained from transcriptions of the training set and serves as the output symbols. For each model, we use a learnable embedding layer to convert symbols to 128-dimensional vectors just before the predictor. The experiments are done using PyTorch \cite{paszke2017automatic} and Fairseq \cite{ott2019fairseq} All models are trained on 32 GPUs with distributed data parallel (DDP) mechanism. We use standard beam search with beam size of 10 for decoding. The decoded sentence pieces are then concatenated into hypotheses to be compared with ground truth transcription for word error rate (WER) evaluation. 

\subsection{Model Architectures and Details}
\label{ssec:architecture}

We compare architectures with roughly the same number of parameters in total. For the encoder in neural transducer, we evaluate options including 1) \texttt{BLSTM 4x640}: bidirectional LSTM with 4 layers of 640 hidden units in each direction, 2) \texttt{LSTM 5x1024}: LSTM with 5 layers of 1024 hidden units and 3) \texttt{Transformer 12x}: VGG-Transformer with 2 layers of VGGNets and 12 Transformer encoder layers. Each VGGNet layer contain 2 layers of two-dimension convolution of 64 kernels of size 3x3. Each Transformer encoder layer takes 512-dimensional inputs, with 8 heads for multi-head self-attention and 2048 as the feed-forward dimension. For efficient inference, all encoders generate output encodings every 60ms. For LSTM/BLSTM this is achieved with low frame rate \cite{pundak2016lower} in which every three consecutive frames are stacked and subsampled to form the new frame, and apply subsampling of factor 2 to the output of the second LSTM/BLSTM layer \cite{he2019streaming}. For VGG-Transformer we set the max-pooling on time dimension to 3 for the 1st VGGNet and 2 for the 2nd VGGNet, as illustrated in Fig. \ref{fig:transformer_encoder_layer}(b).

For the predictor in neural transducer, we evaluate options including 1) \texttt{LSTM 2x700}: LSTM with 2 layers of 700 hidden units and 2) \texttt{Transformer 6x}: VGG-Transformer with 1 layer of VGGNet and 6 Transformer encoder layers. Both the VGGNet layer and the Transformer encoder layers share the same configuration with the the encoder case, with the exception that max-pooling is removed in the VGGNet. In addition, the right context $R$ for these the Transformer encoders is 0 for preventing future information leakage. 

For the joiner in neural transducer, outputs from the encoder $\boldsymbol{h}_{t}$ and the predictor $\boldsymbol{p}_{u}$ are joined with:
\begin{align}
    \boldsymbol{z}_{t, u} = \phi({\boldsymbol{h}_t}W^{h} + {\boldsymbol{p}_{u}}W^{p})W^{o}
\end{align}
where $W^{h} \in \mathcal{R}^{d_{h} \times d_{J}}$ and $W^{p} \in \mathcal{R}^{d_{p} \times d_{J}}$ project $\boldsymbol{h}_{t}$ and $\boldsymbol{p}_{u}$ to a common feature space of dimension $d_{J}$, $\phi()$ is an activation function and $W^{o} \in \mathcal{R}^{d_{J} \times d_{o}}$ generates the logits $\boldsymbol{z}_{t,u}$. We use $d_{J} = 640$, $\phi = ReLU$ and $d_{o} = 256$ consistently for all experiments.

\subsection{Results on Transformer/LSTM Combinations}
\label{ssec:combination_results}

We experimented with combinations of Transformer and LSTM networks for neural transducer. The results are summarized in Table \ref{table:architecture}. For the encoder, we use \texttt{LSTM 5x1024} as the streamable baseline, \texttt{BLSTM 5x640} as the non-streamable baseline and \texttt{Transformer 12x} as the novel replacement for the two. For the predictor, we use \texttt{LSTM 2x700} and \texttt{Transformer 6x} described in section \ref{ssec:architecture} as the two options. 

\begin{table}[htb]
    \centering
    \caption{Neural Transducer with (B)LSTM / Transformer.}
    \scalebox{0.8}{
    \begin{tabular}{|l|l|c|cc|}
    \hline
    \multicolumn{1}{|c|}{\textbf{encoder}} &
    \multicolumn{1}{c|}{\textbf{predictor}} &
    \textbf{\# params} &
    \begin{tabular}{c} \textbf{test-} \\ \textbf{clean} \end{tabular} & 
    \begin{tabular}{c} \textbf{test-} \\ \textbf{other} \end{tabular} \\
    \hline\hline
    (1) LSTM 5x1024 & LSTM 2x700 & 50.5 M & 12.31 & 23.16 \\
    (2) BLSTM 4x640 & LSTM 2x700 & 48.3 M & 6.85 & 16.90 \\
    (3) Transformer 12x & LSTM 2x700 & 45.7 M & \textbf{6.08} & \textbf{13.89} \\
    \hline\hline
    (4) LSTM 5x1024 & Transformer 6x & 67.1 M & 15.76 & 26.67 \\
    (5) BLSTM 4x640 & Transformer 6x & 64.9 M & 7.20 & 16.67 \\
    (6) Transformer 12x & Transformer 6x & 62.3 M & 7.11 & 15.62 \\
    \hline
    \end{tabular}
    }
    \label{table:architecture}
\end{table}

From Table \ref{table:architecture}, given the same configuration for the predictor we see that it is difficult for the LSTM network as encoder to perform well given the constraint on number of parameters. The bidirectional LSTM (BLSTM) network however can compensate the performance and remain compact in size at the cost of being non-streamable. The VGG-Transformer with unlimited self-attention outperforms BLSTM significantly as the encoder and is also non-streamable. For the predictor, for all encoder configurations we see the LSTM network still gives better results than the VGG-Transformer and is smaller in size. As a result we keep \texttt{LSTM 2x700} as the predictor for the experiments in section \ref{ssec:truncated_attention_results}. It is worth noting that the VGG-Transformer loses the advantage of parallel computation as the predictor, as during beam search the hypothesis also extends a token at one search step. 

\subsection{Results on Truncated Self-Attention}
\label{ssec:truncated_attention_results}

We evaluated the impact of the contexts $(L, R)$ in truncated self-attention on recognition accuracy for the VGG-Transformer. As summarized in section \ref{ssec:combination_results}, we find the VGG-Transformer performs well as the encoder but not as the predictor. Therefore we keep \texttt{LSTM 2x700} as the predictor for the experiments in truncated self-attention. The results are summarized in Table \ref{table:truncated}, where $(L, R)$ are used for truncated self-attention in the VGG-Transformer per layer and aggregate through layers. 

\begin{table}[htb]
    \centering
    \caption{Transformer with Truncated Self-Attention.}
    \scalebox{0.8}{
    \begin{tabular}{|l|c|c|cc|}
    \hline
    \multicolumn{1}{|c|}{\textbf{Model Architecture}} &
    $\boldsymbol{L}$ & 
    $\boldsymbol{R}$ & 
    \begin{tabular}{c} \textbf{test-} \\ \textbf{clean} \end{tabular} & 
    \begin{tabular}{c} \textbf{test-} \\ \textbf{other} \end{tabular} \\
    \hline\hline
    (1) LSTM 5x1024 + LSTM 2x700 & inf & 0 & 12.31 & 23.16 \\
    (2) BLSTM 4x640 + LSTM 2x700 & inf & inf & 6.85 & 16.90 \\
    (3) Transformer 12x + LSTM 2x700 & inf & inf & \textbf{6.08} & \textbf{13.89} \\
    \hline\hline
    (4) Transformer 12x + LSTM 2x700 & inf & 0 & 12.32 & 23.08 \\
    (5) Transformer 12x + LSTM 2x700 & inf & 1 & 6.99 & 16.88 \\
    (6) Transformer 12x + LSTM 2x700 & inf & 2 & 6.47 & 15.79 \\
    (7) Transformer 12x + LSTM 2x700 & inf & 4 & \textbf{6.14} & \textbf{14.86} \\
    (8) Transformer 12x + LSTM 2x700 & inf & 8 & 5.99 & 14.17 \\
    \hline\hline
    (9) Transformer 12x + LSTM 2x700 & 4 & 4 & 6.84 & 17.38 \\
    (10) Transformer 12x + LSTM 2x700 & 8 & 4 & 6.69 & 16.79 \\
    (11) Transformer 12x + LSTM 2x700 & 16 & 4 & 6.57 & 15.92 \\
    (12) Transformer 12x + LSTM 2x700 & 32 & 4 & \textbf{6.37} & \textbf{15.30} \\
    \hline
    \end{tabular}
    }
    \label{table:truncated}
\end{table}

Since the right context $R$ introduces algorithmic latency and has major impact on the recognition accuracy, to find optimal parameters for truncated self-attention, we search for the right context $R$ first while keeping the left context $L$ unlimited and then reduce the left context $L$ given the selected right context $R$. From Table \ref{table:truncated} we see both $L$ and $R$ have significant impact on the performance, especially when $R = 0$ when the VGG-Transformer becomes purely causal. However, as $R$ increases, the WERs gradually recover and come close to the case of unlimited self-attention when $R = 8$. With limited right context $R$, the VGG-Transformer becomes streamable but still is $O(T^{2})$ in computational complexity due to the unlimited left context $L$. To keep reasonable performance while minimizing latency at the same time, we selected right context $R = 4$ and evaluate different left contexts $L$. Similar to right context $R$, we see the WER is also sensitive to left context $L$. With $(L, R) = (16, 4)$ we see the VGG-Transformer with truncated self-attention gives better WER than both LSTM/BLSTM baselines. With $(L, R) = (32, 4)$ we only lose 4.7 \% on \texttt{test-clean} and 10.1 \% on \texttt{test-other} relatively compared with the case of umlimited self-attention, but the system becomes streamable and efficient with computational complexity $O(T)$.

\section{Conclusion}
\label{sec:conclusion}

In this paper, we explore options for using the Transformer networks in neural transducer for end-to-end speech recognition. The Transformer network uses self-attention for sequence modeling and can compute in parallel. With causal convolution and truncated self-attention, the neural transducer with the proposed VGG-Transformer as the encoder achieved 6.37 \% on the \texttt{test-clean} set and 15.30 \% on the \texttt{test-other} set of the public corpus LibriSpeech with a small footprint of 45.7 M parameters for the entire system. The proposed Transformer-Transducer is accurate, streamable, compact and efficient, therefore a promising option for resource-limited scenarios such as on-device speech recognition.

\vfill\pagebreak

{
\small
\bibliographystyle{IEEEbib}
\bibliography{strings,refs}
}

\end{document}